\documentclass[aps,preprint,showpacs]{revtex4}%
\usepackage{amsfonts}
\usepackage{amsmath}
\usepackage{amssymb}
\usepackage{graphicx}%
\providecommand{\U}[1]{\protect\rule{.1in}{.1in}}

\begin{document}
\title{Action of the gravitational field on the dynamical Casimir effect}
\author{L. C. C\'{e}leri$^{1,2}$, F. Pascoal$^{1}$, and M. H. Y. Moussa$^{3}$}
\affiliation{$^{1}$Departamento de F\'{\i}sica, Universidade Federal de S\~{a}o Carlos,
Caixa Postal 676, S\~{a}o Carlos, 13565-905, S\~{a}o Paulo,\textit{ }Brazil}
\affiliation{$^{2}$ Universidade Federal do ABC, Centro de Ci\^{e}ncias Naturais e Humanas,
R. Santa Ad\'{e}lia 166, Santo Andr\'{e}, 09210-170, S\~{a}o Paulo, Brasil.}
\affiliation{$^{3}$ Instituto de F\'{\i}sica de S\~{a}o Carlos, Universidade de S\~{a}o
Paulo, Caixa Postal 369, 13560-590 S\~{a}o Carlos, SP, Brazil }

\begin{abstract}
In this paper we analyze the action of the gravitational field on the
dynamical Casimir effect. We consider a massless scalar field confined in a
cuboid cavity placed in a gravitational field described by a static and
diagonal metric. With one of the plane mirrors of the cavity allowed to move,
we compute the average number of particles created inside the cavity by means
of the Bogoliubov coefficients computed through perturbative expansions. We
apply our result to the case of an oscillatory motion of the mirror, assuming
a weak gravitational field described by the Schwarzschild metric. The regime
of parametric amplification is analyzed in detail, demonstrating that our
computed result for the mean number of particles created agrees with specific
associated cases in the literature. Our results, obtained in the framework of
the perturbation theory, are restricted, under resonant conditions, to a
short-time limit.

\end{abstract}

\pacs{03.70.+k; 03.65.Sq; 04.20.Cv; 04.25.Nx.}
\maketitle

\section{Introduction}

Since Casimir's work \cite{Casimir}, we know that vacuum zero-point
fluctuations of a quantum field confined within a finite volume of space exert
radiation pressure on the boundaries that confine the field \cite{Most}. In
fact, the Casimir effect does not require material boundaries; it can be
induced by any classical potential, such as a gravitational field, which is
able to disturb the vacuum, changing the mode structure of this quantum field.
In the case of an electromagnetic field inside a perfectly conducting
Fabri-Perrot cavity, this effect generates an attractive force between the
plates which has been measured with high precision by Lamoureaux \cite{Lamour}
and Mohideen and Roy \cite{Mohideen}. Although Casimir's original analysis
concerned the electromagnetic field, many authors have considered other
fields, for example a fermionic \cite{Queiroz} or the Dirac field
\cite{Chodos}, the latter in the study of the quark confinement problem.

A great deal of effort has been\ devoted to the study of the Casimir effect in
a curved background. The problem of a massless scalar field confined between
two parallel plates placed in a weak and static gravitational field was
considered in Ref. \cite{Sorge}. The author found that the gravitational
interaction causes a small reduction in the Casimir energy, which leads to a
weakening of the force between the plates. In Ref. \cite{Calloni}, the force
acting on a rigid Casimir cavity placed in a weak gravitational field was
computed; it was found that the net force is in the opposite direction from
the gravitational acceleration. An experiment to test the effects of
gravitational curvature on the vacuum energy has also been proposed in Ref.
\cite{Calloni}. Recently, studying the case of an electromagnetic field inside
a plane cavity placed in a weak and static gravitational field, Fulling
\textit{et al.} \cite{Fulling} found that the Casimir energy gravitates as
predicted by the equivalence principle, implying that the virtual field quanta
follow geodesics \cite{Weinberg}.

When the boundaries or, equivalently, the classical potential confining the
field, is time-dependent, the dynamical counterpart of the Casimir effect
takes place, revealing the striking feature of particle creation from the
quantum vacuum. The dynamical Casimir effect (DCE) has been extensively
studied \cite{Dodonov}, since the quantization of the radiation field in a
cavity with moving, perfectly reflecting boundaries, performed by Moore
\cite{Moore} in the early 1970s. However, this phenomenon has not yet been
observed experimentally, despite remarkable efforts \cite{testes}.

The problem of the expanding universe exhibits strong similarities to the DCE
and interesting achievements have been made in that subject
\cite{Parker,Carlitz}. L. Parker \cite{Parker} showed in 1969 that, in an
expanding universe, particles are created from the vacuum \cite{Parker}. In
the same work, it was noted that the initial presence of bosons tends to
increase the number of created bosons, while the opposite is true of fermions.
Working with a brane model for the universe, Durrer \cite{Durrer} showed that
gravitons are formed from vacuum, and Davies \cite{Davies}, studying the
Rindler coordinate system in a flat space-time, found that a uniformly
accelerated observer would see a fixed boundary radiating energy. Actually,
the idea of particle creation due to a nonstatic gravitational field was first
discussed by Schr\"{o}dinger \cite{Schrodinger}, followed by DeWitt
\cite{DeWitt} and Imamura \cite{Imamura}; however, the first who gave a
complete treatment of the problem was Parker \cite{Parker}. It is interesting
to mention that an analogy between the phenomenon of the production of
particles in cosmological models and ion traps has recently been presented
\cite{Ralf}. While gravitons, $\pi$ mesons, protons, and electrons are formed
from the vacuum by the action of the gravitational field, a chain of ions
confined by a time-dependent potential leads to the formation of phonons.

Here we study the action of the gravitational field on the DCE. We consider a
massless scalar field confined in a cuboid cavity placed in a gravitational
field described by a static and diagonal metric. This restriction implies that
the source of the gravitational field does not rotate. One of the plane
mirrors of the cavity is allowed to move in accordance with the law
$a(t)=a_{0}\left(  1+\epsilon f(t)\right)  $, while the remaining boundaries
are fixed; $f(t)$ is an arbitrary function and $\epsilon$ is an exceedingly
small quantity such that $\epsilon\left\vert f(t)\right\vert <<1$. The
restriction $\epsilon\left\vert f(t)\right\vert <<1$ ensures that the
variation of the cavity length in the direction of the gravitational field is
much smaller than its proper length in this direction.

The paper is organized as follows. In Sec. II we perform the quantization of
the scalar field confined inside the cuboid cavity. In Sec. III we compute the
average number of particles created within the cavity, through the well-known
Bogoliubov coefficients \cite{Bogoliubov}. In Sec. IV we apply our general
results to the case of a harmonic motion of \ the mirror, assuming a weak
gravitational field described by the Schwarzschild metric. We emphasize that
our resonant results for the average number of particle creation is restricted
to a short-time limit. We present our concluding remarks in Sec. V. Throughout
this paper we use natural units $c=G=\hbar=1$, adopting the metric signature
$\left(  -,+,+,+\right)  $. We also stress that we refer to the confining
boundary of the scalar field as a \textit{mirror}, because of the assumed
Dirichlet boundary conditions.

\section{Quantization of the massless scalar field}

Let us consider a massless scalar field confined in a closed cuboid cavity,
placed in a gravitational field described by a static and diagonal metric
$g^{\mu\nu}$, with determinant $g$. Dirichlet boundary conditions are imposed
on the scalar field at the mirrors. However, one of the plane mirrors of the
cavity is allowed to move. The equation of motion for the scalar field $\phi$
in the vacuum (where the scalar curvature is null), is given by \cite{Birrel}
\begin{equation}
\partial_{\nu}\left(  \sqrt{-g}g^{\mu\nu}\partial_{\mu}\phi\right)  =0\text{.}
\label{Eq Campo}%
\end{equation}
The Lagrangian density that generates this equation of motion reads%
\[
\mathcal{L}=\frac{1}{2}\sqrt{-g}g^{\mu\nu}\partial_{\mu}\phi\partial_{\nu}%
\phi\text{,}%
\]
with the canonical conjugate momentum defined as%
\begin{equation}
\pi=-\sqrt{-g}g^{00}\partial_{0}\phi\text{.} \label{def mon}%
\end{equation}

Now, we introduce a reference system $\left(  x,y,z\right)  $ with origin on a
static mirror at $z=0$, parallel to the moving one at $z=a(t)$, with
$-\widehat{z}$ denoting the direction of the acceleration of gravity. To
perform the field quantization it is convenient to expand the scalar field in
a complete and orthonormal set of instantaneous mode $\left\{  u_{\mathbf{k}%
}(\mathbf{x};t)\right\}  $ of eigenfrequencies $\omega_{\mathbf{k}}$,
$\mathbf{k=(}k_{x}$, $k_{y}$, $k_{z}\mathbf{)}$ being the associated wave
vector. We thus start with the classical scalar field $\phi(\mathbf{x};t)$ and
its canonical momentum given by
\begin{align}
\phi(\mathbf{x};t)  &  =\sum_{\mathbf{k}}\sqrt{\frac{1}{2\omega_{\mathbf{k}%
}\left(  t\right)  }}\left[  c_{\mathbf{k}}(t)+c_{\mathbf{k}}^{\ast
}(t)\right]  u_{\mathbf{k}}(\mathbf{x};t)\text{,}\label{phi}\\
\pi(\mathbf{x};t)  &  =i\sqrt{-g}g^{00}\sum_{\mathbf{k}}\sqrt{\frac
{\omega_{\mathbf{k}}\left(  t\right)  }{2}}\left[  c_{\mathbf{k}%
}(t)-c_{\mathbf{k}}^{\ast}(t)\right]  u_{\mathbf{k}}(\mathbf{x};t)\text{,}
\label{pi}%
\end{align}
where $c_{\mathbf{k}}(t)$ and $c_{\mathbf{k}}^{\ast}(t)$ are time-dependent
complex coefficients and $\mathbf{x}=\left(  x,y,z\right)  $. The
time-dependence of the instantaneous modes and the\ corresponding
eigenfrequencies, is induced only by the moving mirror \cite{Law,Soff1}. As we
are dealing with a cuboid cavity, we assume that $u_{\mathbf{k}}%
(\mathbf{x};t)$ is a real function that obeys the following differential
equations%
\begin{equation}
\left[  -\sqrt{-g}g^{00}\omega_{\mathbf{k}}^{2}+\partial_{i}\sqrt{-g}%
g^{ij}\partial_{j}\right]  u_{\mathbf{k}}(\mathbf{x};t)=0\text{,}
\label{dif u}%
\end{equation}
normalized by the inner product%
\begin{equation}
-\int_{\mathcal{V}(t)}\operatorname*{d}\nolimits^{3}\mathbf{x}\sqrt{-g}%
g^{00}u_{\mathbf{k}^{\prime}}(\mathbf{x};t)u_{\mathbf{k}}(\mathbf{x}%
;t)=\delta_{\mathbf{kk}^{\prime}}\text{,} \label{Norm}%
\end{equation}
where $i$, $j=x$, $y$, $z$ (from here on) and the integration is performed
over the whole instantaneous cavity volume $\mathcal{V}(t)$. We assume that
the mode functions satisfy the Dirichlet boundary conditions at the mirrors.

The quantization of the scalar field is performed in the canonical form,
constructing a field operator $\Phi$, associated with $\phi$, by promoting the
complex coefficients $c_{\mathbf{k}}$ and $c_{\mathbf{k}}^{\ast}$ to operators
$a_{\mathbf{k}}$ and $a_{\mathbf{k}}^{\dag}$, respectively, and imposing the
equal time commutation relations
\begin{subequations}
\label{RCTI}%
\begin{align}
\left[  \Phi(\mathbf{x};t),\Pi(\mathbf{x}^{\prime};t)\right]   &
=i\delta(\mathbf{x-x}^{\prime})\text{,}\label{RCTIa}\\
\left[  \Pi(\mathbf{x};t),\Pi(\mathbf{x}^{\prime};t)\right]   &  =\left[
\Phi(\mathbf{x};t),\Phi(\mathbf{x}^{\prime};t)\right]  =0\text{,}
\label{RCTIb}%
\end{align}
where $\Pi$ is the field operator associated with $\pi$. As a consequence of
Eqs. (\ref{Norm}) and (\ref{RCTI}), the operators $a_{\mathbf{k}}$ and
$a_{\mathbf{k}}^{\dag}$ satisfy the following commutation relations%
\end{subequations}
\begin{subequations}
\begin{align}
\left[  a_{\mathbf{k}}(t),a_{\mathbf{k}^{\prime}}^{\dag}(t)\right]   &
=\delta_{\mathbf{kk}^{\prime}}\text{,}\label{RCaada}\\
\left[  a_{\mathbf{k}}(t),a_{\mathbf{k}^{\prime}}(t)\right]   &  =\left[
a_{\mathbf{k}}^{\dag}(t),a_{\mathbf{k}^{\prime}}^{\dag}(t)\right]  =0\text{,}
\label{RCaadb}%
\end{align}
which are the usual boson commutation relations for the annihilation and
creation operators. After the definition of these operators, we next compute
the number of particles created inside the cavity by the DCE.

\section{Average number of created particles}

In this section we will compute the mean number of particles created inside
the cavity by the DCE, through the Bogoliubov coefficients $\alpha
_{\mathbf{kk}^{\prime}}(t)$ and $\beta_{\mathbf{kk}^{\prime}}(t)$, defined by
the relations
\end{subequations}
\begin{subequations}
\label{Bog}%
\begin{align}
a_{\mathbf{k}}(t)  &  =\sum_{\mathbf{k}^{\prime}}\left[  \alpha_{\mathbf{kk}%
^{\prime}}(t)a_{\mathbf{k}^{\prime}}(t_{0})+\beta_{\mathbf{kk}^{\prime}%
}(t)a_{\mathbf{k}^{\prime}}^{\dag}(t_{0})\right]  \text{,}\label{Boga}\\
a_{\mathbf{k}}^{\dag}(t)  &  =\sum_{\mathbf{k}^{\prime}}\left[  \alpha
_{\mathbf{kk}^{\prime}}^{\ast}(t)a_{\mathbf{k}^{\prime}}^{\dag}(t_{0}%
)+\beta_{\mathbf{kk}^{\prime}}^{\ast}(t)a_{\mathbf{k}^{\prime}}(t_{0})\right]
\text{,} \label{Bogb}%
\end{align}
which relate the annihilation and creation operators at time $t$, when the
mirror ceases to move, to those at $t=t_{0}$, when the mirror starts to move.
Our strategy is to find, with the help of Eqs. (\ref{phi}) and (\ref{pi}), a
set of differential equations for the operators $a_{\mathbf{k}}$ and
$a_{\mathbf{k}}^{\dag}$. By comparing these equations with the equivalent set
obtained directly from the time derivative of Eq. (\ref{Bog}), we end up with
the desired set of differential equations for the Bogoliubov coefficients.
Expanding these equations in power series of the small parameter $\epsilon$,
we are able to find recurrence relations for both coefficients $\alpha
_{\mathbf{kk}^{\prime}}(t)$ and $\beta_{\mathbf{kk}^{\prime}}(t)$, prompting
their solutions in any desired order.

With the help of Eq. (\ref{Norm}) it is straightforward to obtain, from the
quantum version of Eqs. (\ref{phi}) and (\ref{pi}), the relations
\end{subequations}
\begin{subequations}
\label{aux}%
\begin{align}
a_{\mathbf{k}}(t)  &  =-\sqrt{\frac{\omega_{\mathbf{k}}\left(  t\right)  }{2}%
}\int_{\mathcal{V}(t)}\operatorname*{d}\nolimits^{3}\mathbf{x}\sqrt{-g}%
g^{00}u_{\mathbf{k}}(\mathbf{x};t)\Phi(\mathbf{x};t)\nonumber\\
&  +i\sqrt{\frac{1}{2\omega_{\mathbf{k}}\left(  t\right)  }}\int
_{\mathcal{V}(t)}\operatorname*{d}\nolimits^{3}\mathbf{x}u_{\mathbf{k}%
}(\mathbf{x};t)\Pi(\mathbf{x};t)\text{,}\label{auxa}\\
a_{\mathbf{k}}^{\dag}(t)  &  =-\sqrt{\frac{\omega_{\mathbf{k}}\left(
t\right)  }{2}}\int_{\mathcal{V}(t)}\operatorname*{d}\nolimits^{3}%
\mathbf{x}\sqrt{-g}g^{00}u_{\mathbf{k}}(\mathbf{x};t)\Phi(\mathbf{x}%
;t)\nonumber\\
&  -i\sqrt{\frac{1}{2\omega_{\mathbf{k}}\left(  t\right)  }}\int
_{\mathcal{V}(t)}\operatorname*{d}\nolimits^{3}\mathbf{x}u_{\mathbf{k}%
}(\mathbf{x};t)\Pi(\mathbf{x};t)\text{.} \label{auxb}%
\end{align}
Taking the time derivative of these equations we obtain the above-mentioned
set of differential equations for $a_{\mathbf{k}}$ and $a_{\mathbf{k}}^{\dag}$
as functions of $u_{\mathbf{k}}$, $\Phi$, and $\Pi$, as well as their time
derivatives. Using the relations (\ref{Eq Campo}), (\ref{def mon}), and
(\ref{dif u}), we thus obtain
\end{subequations}
\begin{align}
\dot{a}_{\mathbf{k}}(t)  &  =-i\omega_{\mathbf{k}}a_{\mathbf{k}}%
(t)+\sum_{\mathbf{k}^{\prime}}\left\{  G_{\left[  \mathbf{kk}^{\prime}\right]
}(t)a_{\mathbf{k}^{\prime}}(t)+G_{\left(  \mathbf{kk}^{\prime}\right)
}(t)a_{\mathbf{k}^{\prime}}^{\dag}(t)\right\}  \text{,}\nonumber\\
\dot{a}_{\mathbf{k}}^{\dag}(t)  &  =i\omega_{\mathbf{k}}a_{\mathbf{k}}^{\dag
}(t)+\sum_{\mathbf{k}^{\prime}}\left\{  G_{\left[  \mathbf{kk}^{\prime
}\right]  }(t)a_{\mathbf{k}^{\prime}}^{\dag}(t)+G_{\left(  \mathbf{kk}%
^{\prime}\right)  }(t)a_{\mathbf{k}^{\prime}}(t)\right\}  \text{,}
\label{difa}%
\end{align}
where we have defined the antisymmetric $G_{[\mathbf{kk}^{\prime}%
]}=-G_{[\mathbf{k}^{\prime}\mathbf{k]}}$ and symmetric $G_{(\mathbf{kk}%
^{\prime})}=G_{(\mathbf{k}^{\prime}\mathbf{k)}}$ parts of the coefficients%
\begin{equation}
G_{\mathbf{kk}^{\prime}}(t)=\frac{1}{2}\frac{\dot{\omega}_{\mathbf{k}}}%
{\omega_{\mathbf{k}}}\delta_{\mathbf{kk}^{\prime}}+\sqrt{\frac{\omega
_{\mathbf{k}}}{\omega_{\mathbf{k}^{\prime}}}}\int_{\mathcal{V}(t)}%
\operatorname*{d}\nolimits^{3}\mathbf{x}\sqrt{-g}g^{00}\dot{u}_{\mathbf{k}%
^{\prime}}(\mathbf{x};t)u_{\mathbf{k}}(\mathbf{x};t)\text{.} \label{coef}%
\end{equation}

The equivalent set of equations for $a_{\mathbf{k}}$ and $a_{\mathbf{k}}%
^{\dag}$, generated by differentiating the transformations (\ref{Bog}) is
simply given by
\begin{subequations}
\label{denovo}%
\begin{align}
\dot{a}_{\mathbf{k}}(t)  &  =\sum_{\mathbf{k}^{\prime}}\left[  \dot{\alpha
}_{\mathbf{kk}^{\prime}}(t)a_{\mathbf{k}^{\prime}}(t_{0})+\dot{\beta
}_{\mathbf{kk}^{\prime}}(t)a_{\mathbf{k}^{\prime}}^{\dag}(t_{0})\right]
\text{,}\label{eq1}\\
\dot{a}_{\mathbf{k}}^{\dag}(t)  &  =\sum_{\mathbf{k}^{\prime}}\left[
\dot{\alpha}_{\mathbf{kk}^{\prime}}^{\ast}(t)a_{\mathbf{k}^{\prime}}^{\dag
}(t_{0})+\dot{\beta}_{\mathbf{kk}^{\prime}}^{\ast}(t)a_{\mathbf{k}^{\prime}%
}(t_{0})\right]  \text{.} \label{eq2}%
\end{align}
By substituting the definitions (\ref{Bog}) into Eqs. (\ref{difa}), and
comparing the result with Eqs. (\ref{denovo}), we obtain the set of
differential equations for the Bogoliubov coefficients:
\end{subequations}
\begin{align*}
\dot{\alpha}_{\mathbf{kk}^{\prime}}(t)  &  =-i\omega_{\mathbf{k}}%
\alpha_{\mathbf{kk}^{\prime}}(t)+\sum_{\mathbf{k}^{\prime\prime}}\left\{
G_{\left[  \mathbf{kk}^{\prime\prime}\right]  }(t)\alpha_{\mathbf{k}%
^{\prime\prime}\mathbf{k}^{\prime}}(t)+G_{\left(  \mathbf{kk}^{\prime\prime
}\right)  }(t)\beta_{\mathbf{k}^{\prime\prime}\mathbf{k}^{\prime}}^{\ast
}(t)\right\}  \text{,}\\
\dot{\beta}_{\mathbf{kk}^{\prime}}(t)  &  =-i\omega_{\mathbf{k}}%
\beta_{\mathbf{kk}^{\prime}}(t)+\sum_{\mathbf{k}^{\prime\prime}}\left\{
G_{\left[  \mathbf{kk}^{\prime\prime}\right]  }(t)\beta_{\mathbf{k}%
^{\prime\prime}\mathbf{k}^{\prime}}(t)+G_{\left(  \mathbf{kk}^{\prime\prime
}\right)  }(t)\alpha_{\mathbf{k}^{\prime\prime}\mathbf{k}^{\prime}}^{\ast
}(t)\right\}  \text{.}%
\end{align*}
For reasons of mathematical convenience, we introduce the new coefficients%
\begin{align*}
\tilde{\alpha}_{\mathbf{kk}^{\prime}}  &  =\operatorname*{e}\nolimits^{i\Theta
_{\mathbf{k}}}\alpha_{\mathbf{kk}^{\prime}}\text{,}\\
\tilde{\beta}_{\mathbf{kk}^{\prime}}  &  =\operatorname*{e}\nolimits^{i\Theta
_{\mathbf{k}}}\beta_{\mathbf{kk}^{\prime}}\text{,}%
\end{align*}
leading to the simplified equations
\begin{subequations}
\label{dbog}%
\begin{align}
\frac{\operatorname*{d}\tilde{\alpha}_{\mathbf{kk}^{\prime}}}%
{\operatorname*{d}t}  &  =\sum_{\mathbf{k}^{\prime\prime}}\left[  G_{\left[
\mathbf{kk}^{\prime\prime}\right]  }\operatorname*{e}\nolimits^{i\left[
\Theta_{\mathbf{k}}-\Theta_{\mathbf{k}^{\prime\prime}}\right]  }\tilde{\alpha
}_{\mathbf{k}^{\prime\prime}\mathbf{k}^{\prime}}+G_{\left(  \mathbf{kk}%
^{\prime\prime}\right)  }\operatorname*{e}\nolimits^{i\left[  \Theta
_{\mathbf{k}}+\Theta_{\mathbf{k}^{\prime\prime}}\right]  }\tilde{\beta
}_{\mathbf{k}^{\prime\prime}\mathbf{k}^{\prime}}^{\ast}\right]  \text{,}%
\label{dbog1}\\
\frac{\operatorname*{d}\tilde{\beta}_{\mathbf{kk}^{\prime}}}{\operatorname*{d}%
t}  &  =\sum_{\mathbf{k}^{\prime\prime}}\left[  G_{\left[  \mathbf{kk}%
^{\prime\prime}\right]  }\operatorname*{e}\nolimits^{i\left[  \Theta
_{\mathbf{k}}-\Theta_{\mathbf{k}^{\prime\prime}}\right]  }\tilde{\beta
}_{\mathbf{k}^{\prime\prime}\mathbf{k}^{\prime}}+G_{\left(  \mathbf{kk}%
^{\prime\prime}\right)  }\operatorname*{e}\nolimits^{i\left[  \Theta
_{\mathbf{k}}+\Theta_{\mathbf{k}^{\prime\prime}}\right]  }\tilde{\alpha
}_{\mathbf{k}^{\prime\prime}\mathbf{k}^{\prime}}^{\ast}\right]  \text{,}
\label{dbog2}%
\end{align}
where we have defined $\Theta_{\mathbf{k}}(t)=\int_{t_{0}}^{t}\omega
_{\mathbf{k}}(\tau)\operatorname*{d}\tau$. We have omitted, for notational
simplicity, the explicit time dependence of all parameters. To solve equations
(\ref{dbog}), we expand the\ Bogoliubov coefficients in a power series in
$\epsilon$, written as follows:
\end{subequations}
\begin{subequations}
\label{exp1}%
\begin{align}
\tilde{\alpha}_{\mathbf{kk}^{\prime}}  &  =\sum_{\lambda=0}^{\infty}%
\epsilon^{\lambda}\frac{1}{\lambda!}\lim_{\epsilon\rightarrow0}\frac
{\partial^{\lambda}\tilde{\alpha}_{\mathbf{kk}^{\prime}}}{\partial
\epsilon^{\lambda}}=\sum_{\lambda=0}^{\infty}\epsilon^{\lambda}\tilde{\alpha
}_{\mathbf{kk}^{\prime}}^{(\lambda)}\text{,}\label{exp11}\\
\tilde{\beta}_{\mathbf{kk}^{\prime}}  &  =\sum_{\lambda=0}^{\infty}%
\epsilon^{\lambda}\tilde{\beta}_{\mathbf{kk}^{\prime}}^{(\lambda)}\text{.}
\label{exp22}%
\end{align}
As the coefficients $G_{\mathbf{kk}^{\prime}}$ and $\Theta_{\mathbf{k}}$ also
depend on $\epsilon$ (through $a\left(  t\right)  $), they must be expanded to
compare orders in $\epsilon$ in Eqs. (\ref{dbog}), prompting the relations
\end{subequations}
\begin{subequations}
\label{ep2}%
\begin{align}
G_{\left(  \mathbf{kk}^{\prime}\right)  }\operatorname*{e}\nolimits^{i\left[
\Theta_{\mathbf{k}}+\Theta_{\mathbf{k}^{\prime}}\right]  }  &  =\sum
_{\lambda=1}^{\infty}\epsilon^{\lambda}\Lambda_{\mathbf{kk}^{\prime}%
}^{(\lambda)}\text{,}\label{ep11}\\
G_{\left[  \mathbf{kk}^{\prime}\right]  }\operatorname*{e}\nolimits^{i\left[
\Theta_{\mathbf{k}}-\Theta_{\mathbf{k}^{\prime}}\right]  }  &  =\sum
_{\lambda=1}^{\infty}\epsilon^{\lambda}\Xi_{\mathbf{kk}^{\prime}}^{(\lambda
)}\text{.} \label{ep22}%
\end{align}
The sums in the above expansions start with $\lambda=1$ because the
coefficients $G_{\mathbf{kk}^{\prime}}$ are proportional to $\dot{a}$ and,
consequently, their lowest contribution is of first order, as expected.

By substituting Eqs. (\ref{exp1}) and (\ref{ep2}) into Eq. (\ref{dbog}) and
comparing the terms of the same order in $\epsilon$ we find, with the help of
the initial conditions $\tilde{\alpha}_{\mathbf{kk}^{\prime}}^{(0)}%
(t)=\delta_{\mathbf{kk}^{\prime}}$ and $\tilde{\beta}_{\mathbf{kk}^{\prime}%
}^{(0)}(t)=0$, the recurrence relations
\end{subequations}
\begin{subequations}
\label{recnovo}%
\begin{align}
\tilde{\alpha}_{\mathbf{kk}^{\prime}}^{(\lambda)}(t)  &  =\sum_{\mathbf{k}%
^{\prime\prime}}\sum_{\lambda^{\prime}=0}^{\lambda-1}\int_{t_{0}}%
^{t}\operatorname*{d}\tau\left\{  \Xi_{\mathbf{kk}^{\prime\prime}}%
^{(\lambda-\lambda^{\prime})}(\tau)\tilde{\alpha}_{\mathbf{k}^{\prime\prime
}\mathbf{k}^{\prime}}^{(\lambda^{\prime})}(\tau)\right. \nonumber\\
&  \left.  +\Lambda_{\mathbf{kk}^{\prime\prime}}^{(\lambda-\lambda^{\prime}%
)}(\tau)\tilde{\beta}_{\mathbf{k}^{\prime\prime}\mathbf{k}^{\prime}}%
^{(\lambda^{\prime})\ast}(\tau)\right\}  \text{,}\label{rec1}\\
\tilde{\beta}_{\mathbf{kk}^{\prime}}^{(\lambda)}(t)  &  =\sum_{\mathbf{k}%
^{\prime\prime}}\sum_{\lambda^{\prime}=0}^{\lambda-1}\int_{t_{0}}%
^{t}\operatorname*{d}\tau\left\{  \Xi_{\mathbf{kk}^{\prime\prime}}%
^{(\lambda-\lambda^{\prime})}(\tau)\tilde{\beta}_{\mathbf{k}^{\prime\prime
}\mathbf{k}^{\prime}}^{(\lambda^{\prime})}(\tau)\right. \nonumber\\
&  \left.  +\Lambda_{\mathbf{kk}^{\prime\prime}}^{(\lambda-\lambda^{\prime}%
)}(\tau)\tilde{\alpha}_{\mathbf{k}^{\prime\prime}\mathbf{k}^{\prime}%
}^{(\lambda^{\prime})\ast}(\tau)\right\}  \text{,} \label{rec2}%
\end{align}
which provide the Bogoliubov coefficients to any desired order of parameter
$\epsilon$ and, consequently, the average number of particles created inside
the cavity. Assuming that the scalar field is initially in the vacuum state
$\left\vert \left\{  0_{\mathbf{k}}\right\}  \right\rangle =%
{\textstyle\prod\nolimits_{\mathbf{k}}}
\left\vert 0_{\mathbf{k}}\right\rangle $, where $a_{\mathbf{k}}(t_{0}%
)\left\vert \left\{  0_{\mathbf{k}}\right\}  \right\rangle =0$, the average
number of particles created in the $\mathbf{k}$th mode, computed from Eqs.
(\ref{Bog}), is given by
\end{subequations}
\begin{align}
\mathcal{N}_{\mathbf{k}}(t)  &  =\left\langle \left\{  0_{\mathbf{k}^{\prime}%
}\right\}  \right\vert a_{\mathbf{k}}^{\dag}(t)a_{\mathbf{k}}(t)\left\vert
\left\{  0_{\mathbf{k}^{\prime}}\right\}  \right\rangle \nonumber\\
&  =\sum_{\mathbf{k}^{\prime}}\left\vert \beta_{\mathbf{k}^{\prime}\mathbf{k}%
}(t)\right\vert ^{2}=\sum_{\mathbf{k}^{\prime}}\left\vert \tilde{\beta
}_{\mathbf{k}^{\prime}\mathbf{k}}(t)\right\vert ^{2}\text{.} \label{Num}%
\end{align}

In the next section we apply this result to the particular case of an
oscillatory motion of the mirror and a Schwarzschild background.

\section{Oscillatory motion of the mirror}

Let us consider a cavity of dimensions $a_{0}\times a_{0}\times a(t)$, $a(t)$
specifying the sinoidal law of motion
\[
a(t)=a_{0}\left[  1+\epsilon\sin(\varpi t)\right]  \text{,}%
\]
where $\varpi$ is the oscillation frequency of the mirror and $\epsilon\ll1$.
We also assume that the cavity is placed at a distance $R$ from the
gravitational source (outside the mass distribution), represented by a
nonrotating spherical mass $M$. The gravitational field is thus described by
the Schwarzschild metric which, in the isotropic coordinates and weak-field
limit $M/r\ll1$ --- $r$ being the radial coordinate --- is given by
\cite{Birrel}%
\begin{equation}
\text{d}s^{2}=-\left(  1-2\frac{M}{r}\right)  \text{d}t^{2}+\left(
1+2\frac{M}{r}\right)  \text{d}\mathbf{r}^{2}\text{.} \label{1}%
\end{equation}

Under the realistic restriction that the dimensions of the cavity are
negligible compared to the size of the gravitational source, i. e., $a_{0}\ll
R$, we expand the line element (\ref{1}) over a short distance $z$ around $R$
in the radial direction \cite{Sorge, Fulling} $r^{2}=x^{2}+y^{2}+\left(
z+R\right)  ^{2}$ up to order $(M/R)^{2}$, to obtain
\begin{equation}
\frac{M}{r}\simeq\chi-\gamma z\text{, } \label{exp}%
\end{equation}
where $\chi=M/R$ and $\gamma=M/R^{2}$ is the acceleration of gravity. With
this expansion the line element (\ref{1}) can be rewritten as%
\begin{equation}
\text{d}s^{2}=-\left(  1-2\chi+2\gamma z\right)  \text{d}t^{2}+\left(
1+2\chi-2\gamma z\right)  \text{d}\mathbf{r}^{2}\text{.} \label{3}%
\end{equation}

Next, we consider only the first order approximation, where $\gamma=0$.

\subsection{First-order approximation: a constant gravitational field}

Considering the first-order approximation in the expansion (\ref{exp}), i.e.,
$\gamma=0$, the instantaneous mode functions --- that satisfy Eqs.
(\ref{dif u}) and (\ref{Norm}), as well as the Dirichlet boundary conditions
at the mirrors--- are given by
\begin{equation}
u_{\mathbf{k}}(\mathbf{x};t)=\frac{2\left(  1-2\chi\right)  }{a_{0}}%
\sin\left[  k_{x}(0)x\right]  \sin\left[  k_{y}(0)y\right]  \sqrt{\frac
{2}{a(t)}}\sin\left[  k_{z}(t)z\right]  \text{,} \label{u1}%
\end{equation}
where $k_{i}(t)=\left(  n_{i}\pi\right)  /a(t)$ and $n_{i}$ stands for the
positive integers. The corresponding eigenfrequencies read%
\begin{equation}
\omega_{\mathbf{k}}(t)=\left(  1-2\chi\right)  \sqrt{k_{x}^{2}(0)+k_{y}%
^{2}(0)+k_{z}^{2}(t)}\text{.} \label{freq1}%
\end{equation}

As we are assuming that $\epsilon$ is a small number, we will compute the mean
number of particles created up to second order in this parameter, which is the
first non-zero contribution. This implies, as we can see from Eq. (\ref{Num}),
that we have to compute the coefficient $\tilde{\beta}_{\mathbf{kk}^{\prime}}$
up to first order in $\epsilon$, given the result%

\begin{equation}
\mathcal{N}_{\mathbf{k}}(t)\simeq\epsilon^{2}\sum_{\mathbf{k}^{\prime}%
}\left\vert \tilde{\beta}_{\mathbf{kk}^{\prime}}^{(1)}(t)\right\vert
^{2}=\epsilon^{2}\sum_{\mathbf{k}^{\prime}}\left\vert \int_{t_{0}}%
^{t}\operatorname*{d}\tau\Lambda_{\mathbf{kk}^{\prime}}^{(1)}(\tau)\right\vert
^{2}\text{,} \label{num1}%
\end{equation}
where we have used Eq. (\ref{recnovo}). With the definition $\mathbf{n}%
^{2}=n_{x}^{2}+n_{y}^{2}+n_{z}^{2}$, the first-order approximation of the
coupling coefficients $G_{\mathbf{kk}^{\prime}}(t)$ in Eq. \ref{ep11} read%
\begin{align*}
\Lambda_{\mathbf{kk}^{\prime}}^{(1)}(t)  &  =\delta_{n_{x},n_{x}^{\prime}%
}\delta_{n_{y},n_{y}^{\prime}}\varpi\cos(\varpi t)\operatorname*{e}%
\nolimits^{i\omega_{n_{z},n_{z}^{\prime}}t}\left\{  -\delta_{n_{z}%
,n_{z}^{\prime}}\frac{n_{z}^{2}}{2\mathbf{n}^{2}}\right. \\
&  \left.  +(1-\delta_{n_{z},n_{z}^{\prime}})(-1)^{n_{z}+n_{z}^{\prime}}%
\frac{n_{z}n_{z}^{\prime}}{n_{z}^{\prime2}-n_{z}^{2}}\frac{\mathbf{n}%
-\mathbf{n}^{\prime}}{\sqrt{\mathbf{n}\text{ }\mathbf{n}^{\prime}}}\right\}
\text{,}%
\end{align*}
where we have defined the frequencies%
\[
\omega_{\mathbf{n},\mathbf{n}^{\prime}}=\delta_{n_{x},n_{x}^{\prime}}%
\delta_{n_{y},n_{y}^{\prime}}\left(  1-2\chi\right)  \frac{\pi}{a_{0}}\left(
\mathbf{n}+\mathbf{n}^{\prime}\right)  \text{.}%
\]
By substituting the expressions for $\Lambda_{\mathbf{kk}^{\prime}}^{(1)}(t)$
and $\omega_{\mathbf{n},\mathbf{n}^{\prime}}$ into Eq. (\ref{num1}), we obtain
the following expression for the mean number of particles created in a
selected mode $\mathbf{k}$%
\[
\mathcal{N}_{\mathbf{k}}=\frac{1}{4}\sum_{\mathbf{n}^{\prime}}\epsilon
^{2}\varpi^{2}t^{2}\mathcal{C}_{\mathbf{n},\mathbf{n}^{\prime}}\left\vert
f_{\mathbf{n},\mathbf{n}^{\prime}}(\varpi,t)\right\vert ^{2}\text{,}%
\]
where the constant coupling coefficients $\mathcal{C}_{\mathbf{n}%
,\mathbf{n}^{\prime}}$ are
\[
\mathcal{C}_{\mathbf{n},\mathbf{n}^{\prime}}=\delta_{n_{x},n_{x}^{\prime}%
}\delta_{n_{y},n_{y}^{\prime}}\left\{  \frac{1}{4}\delta_{n_{z},n_{z}^{\prime
}}\left(  \frac{n_{z}^{2}}{\mathbf{n}^{2}}\right)  ^{2}+(1-\delta_{n_{z}%
,n_{z}^{\prime}})\frac{n_{z}^{2}n_{z}^{2\prime}}{\left(  n_{z}^{\prime2}%
-n_{z}^{2}\right)  ^{2}}\frac{\left(  \mathbf{n}-\mathbf{n}^{\prime}\right)
^{2}}{\mathbf{n}\text{ }\mathbf{n}^{\prime}}\right\}  \text{,}%
\]
and the time-dependent function $f_{\mathbf{n},\mathbf{n}^{\prime}}$ is given
by
\begin{equation}
f_{\mathbf{n},\mathbf{n}^{\prime}}(\varpi,t)=\delta_{n_{x},n_{x}^{\prime}%
}\delta_{n_{y},n_{y}^{\prime}}\left\{  \frac{\exp\left[  i\left(
\omega_{\mathbf{n},\mathbf{n}^{\prime}}-\varpi\right)  t\right]  -1}{\left(
\omega_{\mathbf{n},\mathbf{n}^{\prime}}-\varpi\right)  t}+\frac{\exp\left[
i\left(  \omega_{\mathbf{n},\mathbf{n}^{\prime}}+\varpi\right)  t\right]
-1}{\left(  \omega_{\mathbf{n},\mathbf{n}^{\prime}}+\varpi\right)  t}\right\}
\text{.} \label{fcoef}%
\end{equation}
As we can see from Eq. (\ref{fcoef}), $\mathcal{N}_{\mathbf{k}}$ is an
oscillatory function in time, except when at least one of the resonance
conditions $\varpi=\omega_{\mathbf{n},\mathbf{n}^{\prime}}$ is satisfied.
(Note that the second high-oscillatory term in the RHS of Eq. (\ref{fcoef})
can be disregarded within the rotating wave approximation.) Therefore, under
the resonance condition, the mean number of particles created is a function
that increases quadratically in time, given by%
\[
\lim_{\varpi\rightarrow\omega_{\mathbf{n},\mathbf{n}^{\prime}}}\mathcal{N}%
_{\mathbf{k}}\simeq\frac{1}{4}\mathcal{C}_{\mathbf{n},\mathbf{n}^{\prime}%
}\left(  \epsilon\omega_{\mathbf{n},\mathbf{n}^{\prime}}t\right)  ^{2}\text{.}%
\]
We remember that $r$ and $t$ are only coordinates, without any direct physical
significance (see the discussions in Refs. \cite{Sorge, Fulling}). To obtain a
measurable quantity, we must rewrite this result in terms of the proper time
and length of the cuboid cavity, defined, in this static case, as $t_{p}=\int
dt\sqrt{-g_{00}}$ and $a_{p}=\int dz\sqrt{g_{zz}}$, respectively. With this
consideration we obtain the quantity%
\begin{equation}
\omega_{\mathbf{n},\mathbf{n}^{\prime}}t=\delta_{n_{x},n_{x}^{\prime}}%
\delta_{n_{y},n_{y}^{\prime}}\frac{\pi}{a_{p}}\left(  \mathbf{n}%
+\mathbf{n}^{\prime}\right)  t_{p}\text{,} \label{omega}%
\end{equation}
given, under the resonance condition, the expected mean number of particles
created due to the DCE
\[
\lim_{\varpi\rightarrow\omega_{\mathbf{n},\mathbf{n}^{\prime}}}\mathcal{N}%
_{\mathbf{k}}\simeq\delta_{n_{x},n_{x}^{\prime}}\delta_{n_{y},n_{y}^{\prime}%
}\mathcal{C}_{\mathbf{n},\mathbf{n}^{\prime}}\left(  \epsilon\frac{\pi}%
{2a_{p}}\left(  \mathbf{n}+\mathbf{n}^{\prime}\right)  t_{p}\right)
^{2}\text{.}%
\]

The above first-order result shows that a constant gravitational field does
not modify the number of particles created due to the DCE inside the cavity
(note that the coefficients $\mathcal{C}_{\mathbf{n},\mathbf{n}^{\prime}}$ do
not depend on $t$ or $\omega_{\mathbf{n},\mathbf{n}^{\prime}}$). The fact that
a constant field does not change the static Casimir energy is a consequence of
the lack of dependence of the physical results on the origin of the coordinate
system \cite{Sorge, Fulling}. Therefore, any contribution of the gravitational
field to the number of particles created will appear at least in a
second-order approximation, where the spatial dependence of the metric arises.
Next, we compute such a second-order correction for the number of particles.

\subsection{Second-order approximation}

In this section, considering the full metric shown in Eq. (\ref{3}), we seek
mode solutions in the form%
\begin{equation}
u_{\mathbf{k}}(\mathbf{x};t)=\frac{2}{a_{0}}\sin\left[  k_{x}(0)x\right]
\sin\left[  k_{y}(0)y\right]  \xi_{\mathbf{k}}(z;t)\text{.} \label{mode1}%
\end{equation}
By substituting this ansatz solution into Eq. (\ref{dif u}), the following
differential equation for $\xi_{\mathbf{k}}$ results:%
\begin{equation}
\partial_{z}^{2}\xi_{\mathbf{k}}-4\gamma\omega_{\mathbf{k}}^{2}z\xi
_{\mathbf{k}}=-\Omega_{\mathbf{k}}^{2}\xi_{\mathbf{k}}\text{,} \label{9}%
\end{equation}
where we have defined $\Omega_{\mathbf{k}}^{2}=\omega_{\mathbf{k}}^{2}\left(
1+4\chi\right)  -\left(  n_{x}\pi/a_{0}\right)  ^{2}-\left(  n_{y}\pi
/a_{0}\right)  ^{2}$\textbf{.}

Now, following the reasoning in Ref. \cite{Sorge}, it is convenient to perform
the coordinate transformation
\begin{equation}
v_{\mathbf{k}}(z)=\left(  \frac{\Omega_{\mathbf{k}}^{2}}{4\gamma
\omega_{\mathbf{k}}^{2}}-z\right)  \left(  4\gamma\omega_{\mathbf{k}}%
^{2}\right)  ^{1/3}\text{,} \label{10}%
\end{equation}
which leads to the following simplified form of the Airy differential equation%
\[
\partial_{v}^{2}\xi_{\mathbf{k}}(v_{\mathbf{k}})+v\xi_{\mathbf{k}%
}(v_{\mathbf{k}})=0\text{,}%
\]
whose solutions can be written in terms of a linear combination of Bessel
functions of the first kind:%
\[
\xi_{\mathbf{k}}(v_{\mathbf{k}})=\sqrt{v_{\mathbf{k}}}\left[  A_{\mathbf{k}%
}J_{1/3}\left(  \frac{2}{3}v_{\mathbf{k}}^{3/2}\right)  +B_{\mathbf{k}%
}J_{-1/3}\left(  \frac{2}{3}v_{\mathbf{k}}^{3/2}\right)  \right]  \text{.}%
\]
By applying the boundary conditions and noting, from Eq. (\ref{10}), that
$v_{\mathbf{k}}(z,t)>>1$ for all values of $t$, we obtain the approximate
solution \cite{Sorge}%
\begin{equation}
\xi_{\mathbf{k}}(v_{\mathbf{k}})\simeq N_{\mathbf{k}}v_{\mathbf{k}}%
^{-1/4}(z)\sin\left(  \frac{2}{3}v_{\mathbf{k}}^{3/2}(z)-\frac{2}%
{3}v_{\mathbf{k}}^{3/2}(0)\right)  \text{,} \label{mode2}%
\end{equation}
with the normalization factor $N_{\mathbf{k}}$ being fixed by Eq.
(\ref{Norm}), and the expression
\begin{equation}
\omega_{\mathbf{k}}(t)\simeq\left[  1-2\chi+\gamma a(t)\right]  \sqrt
{k_{x}^{2}(0)+k_{y}^{2}(0)+k_{z}^{2}(t)}\text{,} \label{11}%
\end{equation}
corroborating the above first-order approximation for $\omega_{\mathbf{k}}(t)$.

Therefore, by computing the coefficients $\Lambda_{\mathbf{kk}^{\prime}}%
^{(1)}$ through the mode functions written in Eqs. (\ref{mode1}) and
(\ref{mode2}), we finally obtain the second-order result for the mean number
of particles created in mode $\mathbf{k}$, given by%

\begin{equation}
\mathcal{N}_{\mathbf{k}}=\frac{1}{4}\sum_{\mathbf{n}^{\prime}}\epsilon
^{2}\varpi^{2}t^{2}\mathcal{C}_{\mathbf{n},\mathbf{n}^{\prime}}^{(2)}%
\left\vert f_{\mathbf{n},\mathbf{n}^{\prime}}(\varpi,t)\right\vert
^{2}\text{,} \label{Nk}%
\end{equation}
where the function $f_{\mathbf{n},\mathbf{n}^{\prime}}$ is defined in Eq.
(\ref{fcoef}), but with the frequencies $\omega_{\mathbf{n},\mathbf{n}%
^{\prime}}$ derived from Eq. (\ref{11}) as%
\[
\omega_{\mathbf{n},\mathbf{n}^{\prime}}=\delta_{n_{x},n_{x}^{\prime}}%
\delta_{n_{y},n_{y}^{\prime}}\left(  1-2\chi+\gamma a_{0}\right)  \frac{\pi
}{a_{0}}\left(  \mathbf{n}+\mathbf{n}^{\prime}\right)  \text{.}%
\]

A general expression for the coefficient $\mathcal{C}_{\mathbf{n}%
,\mathbf{n}^{\prime}}^{(2)}$ (valid for any value of the mirror frequency
$\varpi$), can be computed with the help of Eqs. (\ref{mode1}) and
(\ref{mode2}). Its solution, though too cumbersome to be shown here, is
straightforward. Therefore, we consider a special case of interest, the one
that maximizes the number of particles created in a selected mode $\mathbf{k}%
$: the parametric amplification process, in which the frequency of the mirror
oscillation is twice the stationary eigenfrequency of a given mode of the
cavity, i. e., $\varpi=2\omega_{\mathbf{k}}(0)$. As seen from Eqs.
(\ref{fcoef}) and (\ref{omega}), this condition implies that $\mathbf{k=k}%
^{\prime}$. Under these assumptions, the coefficients $\mathcal{C}%
_{\mathbf{n},\mathbf{n}}^{(2)}$ have the simple expression%
\begin{equation}
\mathcal{C}_{\mathbf{n},\mathbf{n}}^{(2)}\simeq\frac{1}{4}\left(  \frac
{n_{z}^{2}}{\mathbf{n}^{2}}-\gamma a_{0}\right)  ^{2}\text{,} \label{coef2}%
\end{equation}
so that the number of particles created in mode $\mathbf{k}$ is given by%
\[
\mathcal{N}_{\mathbf{k}}\simeq\left(  \frac{\epsilon\omega_{\mathbf{k}}%
(0)t}{2}\right)  ^{2}\left(  \gamma a_{0}-\frac{n_{z}^{2}}{\mathbf{n}^{2}%
}\right)  ^{2}\text{.}%
\]
Rewriting this equation in terms of the proper length and time%
\begin{align*}
a_{0}  &  \simeq a_{p}\left(  1+\chi+\gamma\frac{a_{p}}{2}\right)  \text{,}\\
t  &  \simeq\left(  1+\chi-\gamma a_{p}\right)  t_{p}\text{,}%
\end{align*}
we obtain our final result%
\begin{equation}
\mathcal{N}_{\mathbf{k}}=\left[  \frac{n_{z}^{2}}{\mathbf{n}^{2}}\left(
1-4\chi\right)  -\gamma a_{p}\left(  1+\frac{n_{z}^{2}}{\mathbf{n}^{2}%
}\right)  \right]  ^{2}\left(  \mathbf{n}\tau_{p}\right)  ^{2}\text{,}
\label{final}%
\end{equation}
where we have defined the dimensionless proper time variable $\tau
_{p}=\epsilon\pi t_{p}/2a_{0}$. In the absence of gravity, Eq. (\ref{final})
simplifies to $\mathcal{N}_{\mathbf{k}}=\left(  n_{z}/\mathbf{n}\right)
^{4}\left(  \mathbf{n}\tau_{p}\right)  ^{2}$, recovering the result in the
literature for the DCE under the parametric amplification condition. For the
fundamental mode, i. e., $n_{x}=n_{y}=n_{z}=1$, we obtain%
\begin{equation}
\mathcal{N}_{\mathbf{1}}=\left[  1-4\chi-2\gamma a_{p}\right]  ^{2}\tau
_{p}^{2}\text{,} \label{N1}%
\end{equation}
instead of the simpler result $\mathcal{N}_{\mathbf{1}}=\tau_{p}^{2}$ given by
the DCE in a flat space-time or, as demonstrated above, within the first-order
approximation $\gamma=0$. This expression shows that the effect of the
gravitational field on the DCE is simply to diminish the number of created
particles. This fact is in agreement with the results obtained in Refs.
\cite{Sorge, Fulling}, where the authors demonstrate that the coupling with a
gravitational field causes a weakening of the Casimir force.

Although we have considered only the parametric resonance regime --- creating
degenerate pairs of particles --- other resonances can evidently be satisfied
by the moving mirror: apart from the nondegenerate creation of pairs of
particles in distinct modes $\mathbf{k}$ and $\mathbf{k}^{\prime}$, under the
resonance condition $\varpi=\omega_{\mathbf{k}}+\omega_{\mathbf{k}^{\prime}}$,
the scattering of particles between these modes also takes place under the
condition $\varpi=\left\vert \omega_{\mathbf{k}}-\omega_{\mathbf{k}^{\prime}%
}\right\vert $. However, in a three-dimensional cavity, a suitable choice of
the cavity dimensions forbids both degenerate and nondegenerate resonances
from occurring simultaneously. Therefore, it is reasonable to consider only
the degenerate amplification process when analyzing the particle creation
mechanism by the DCE under the action of a gravitational field \cite{Soff2}.
Evidently, the mean number of particles created can be computed under any
resonance condition from Eq. (\ref{Nk}), by performing a numerical calculation
of the second-order coefficients $\mathcal{C}_{\mathbf{n},\mathbf{n}^{\prime}%
}^{(2)}$.

In Fig. 1, we plot $\mathcal{N}_{\mathbf{1}}$ against the dimensionless
parameter $\gamma a_{p}$ for a fixed value $\tau_{p}=0.1$ and $\chi=0$. As
expected from Eq. (\ref{N1}), the number of particles created in the
fundamental mode decreases as $\gamma$ increases. In Fig. 2, we plot
$\mathcal{N}_{\mathbf{k}}$ against $\gamma a_{p}$ for a fixed value $\tau
_{p}=0.1/\mathbf{n}$ (during which the cavity performs $1/2\pi\epsilon$
oscillations), and for a few values of $\mathbf{n}$. Again as expected, since
the energy of a given mode increases with $\mathbf{n}$, the number of
particles created decreases as $\mathbf{n}$ increases.

\section{Concluding remarks}

In this paper we analyzed the action of the gravitational field on the number
of particles created in a massless scalar field by the dynamical Casimir
effect. We considered a cuboid cavity, with one of its plane mirrors allowed
to move, placed in a static gravitational field described by a diagonal
metric. We first computed the mean number of particles created under an
arbitrary law of motion of the cavity mirror, employing the Bogoliubov
coefficients obtained by perturbative expansions. Next, the mean number is
analyzed under the particular circumstances of an oscillatory motion of the
mirror and a weak gravitational field described by the Schwarzschild metric.
As already emphasized, our resonant results are restricted to the short-time
approximation $\tau_{p}\ll1$.

Our first-order result, that a constant gravitational field does not affect
the mean number of particles created by the DCE, is in agreement with those in
Refs. \cite{Sorge, Fulling} showing that a constant field does not change the
static Casimir energy. The reason for this behavior is the fact that the
physical results do not depend on the origin of the coordinate system
\cite{Sorge, Fulling}. Therefore, the effetc of the gravitational field
appears only in a second or higher order approximation, in which the spatial
dependence of the metric arises.

Considering only the parametric resonance regime, our second-order result
shows that the mean number of created particles is diminished by the
gravitational field, again in agreement with the fact that the coupling with
this field weakens the Casimir force \cite{Sorge, Fulling}. Expressed
differently, as the frequencies of the cavity scalar field are red-shifted
relative to their values in the absence of gravity, the mean number of created
particles, proportional to the square of these frequencies, must decrease.

We observe that the effect of the temperature, which is imporatant for the
experimental verification of the DCE, can be taken into account by considering
a non-null scalar curvature in the cavity field equation. Moreover, the
equivalence principle prevents the orientation of the Casimir apparatus (with
respect to the gravity acceleration) contributing to the DCE \cite{Fulling}.

We finally recall from the Introduction that the possibility of the
experimental observation of the force produced by vacuum fluctuations acting
on a rigid Casimir cavity in a weak gravitational field was discussed in Ref.
\cite{Calloni}. Although such an experimental test of the effects reported
here is certainly no less challenging than the verification of the DCE itself,
the action of gravity on the quantum vacuum fluctuation is a fundamental issue
which could play a significant role on the evolution the universe.

\textbf{Acknowledgements}

We wish to express thanks for the support from CNPq and FAPESP, Brazilian agencies.


\begin{thebibliography}{99}                                                                                               %


\bibitem {Casimir}H. B. G. Casimir, Proc. K. Ned. Akad. Wet. \textbf{51}, 793 (1948).

\bibitem {Most}M. Bordag, U. Mohideen, and V. M. Mostepanenko, Phys. Rep.
\textbf{353}, 1 (2001); K. A. Milton, \textit{The Casimir Effect: Physical
Manifestations of Zero-Point Energy} (World Scientific, Singapore, 2001).

\bibitem {Lamour}S. K. Lamoreaux, Phys. Rev. Lett. \textbf{78}, 5 (1997).

\bibitem {Mohideen}U. Mohideen and A. Roy, Phys. Rev. Lett. \textbf{81}, 4549 (1998).

\bibitem {Queiroz}H. Queiroz, J.C. da Silva, F.C. Khanna, J. M. C.
Malbouisson, M. Revzen, A. E. Santana, Ann. Phys. \textbf{317}, 220 (2005).

\bibitem {Chodos}A. Chodos, R. L. Jaffe, K. Johnson, and C. B. Thorn, Phys.
Rev. D \textbf{10}, 2599 (1974).

\bibitem {Sorge}F. Sorge, Class. Quantum Grav. \textbf{22}, 5109 (2005);

\bibitem {Calloni}E. Calloni, L. Di Fiore, G. Esposito, L. Milano, and L.
Rosa, Phys. Lett. A \textbf{297}, 328 (2002); E. Calloni, L. Di Fiore, G.
Esposito, L. Milano, and L. Rosa, Int. J. Mod. Phys. A \textbf{17}, 804 (2002).

\bibitem {Fulling}S. A. Fulling, K. A. Milton, P. Parashar, A. Romeo, K. V.
Shajesh, and J. Wagner, Phys. Rev. D \textbf{76}, 025004 (2007);K. A. Milton,
S. A. Fulling, P. Parashar, A. Romeo, K. V. Shajesh, and J. Wagner, J. Phys.
A: Math. Theor. \textbf{41}, 164052 (2008).

\bibitem {Weinberg}S. Weinberg, \textit{Gravitation and Cosmology: Principles
and Applications of the Theory of General Relativity} (Wiley, New York, 1972).

\bibitem {Dodonov}V. V. Dodonov, in \textit{Modern Nonlinear Optics}, Part 3,
edited by M. W. Evans, Adv. Chem. Phys. Series \textbf{119} (Wiley, New York,
2001), p. 309; G. Barton and C. Eberlein, Ann. Phys. (Berlin) \textbf{227},
222 (1993).

\bibitem {Moore}G. T. Moore, J. Math. Phys. (N.Y.) \textbf{11}, 2679 (1970).

\bibitem {testes}C. Braggio, G. Bressi, G. Carugno, C. Del Noce, G. Galeazzi,
A. Lombardi, A. Palmieri, G. Ruoso, and D. Zanello, Eurphysics Lett.
\textbf{70}, 754 (2005); A. Agnesi, C. Braggio, G. Bressi, G. Carugno, G.
Galeazzi, F. Pirzio, G. Reali, G. Ruoso, and D, Zanello, J. Phys. A: Math.
Theor. \textbf{41}, 164024 (2008); W. -J. Kim, J. H. Brownell, and R. Onofrio,
Phys. Rev. Lett. \textbf{96}, 200402 (2006); A. Lambrecht, M. -T. Jaekel, and
S. Reynaud, Phys. Rev. Lett. \textbf{77}, 615 (1996).

\bibitem {Parker}L. Parker, Phys. Rev. Lett. \textbf{21}, 562 (1968); Phys.
Rev. \textbf{183}, 1057 (1969); Phys. Rev. D \textbf{3}, 346 (1970).

\bibitem {Carlitz}R. D. Carlitz and R. S. Willey, Phys. Rev. D \textbf{36},
2327 (1987); P. C. W. Davies, J. Opt. B: Quantum Semiclass. Opt. \textbf{7},
S40 (2005).

\bibitem {Durrer}R. Durrer and M. Ruser, Phys. Rev. Lett. \textbf{99}, 071601 (2007).

\bibitem {Davies}P. C. W. Davies, J. Phys. A: Math. Gen. \textbf{8}, 609 (1975).

\bibitem {Schrodinger}E. Schrodinger, Physica (Utrecht) \textbf{6}, 899 (1939).

\bibitem {DeWitt}B. S. DeWitt, Phys. Rev. \textbf{90}, 357 (1953).

\bibitem {Imamura}T. Imamura, Phys. Rev. \textbf{118}, 1430 (1960).

\bibitem {Ralf}R. Sch\"{u}tzhold, M. Uhlmann, L. Petersen, H. Schmitz, A.
Friedenauer, and T. Sch\"{a}tz, Phys. Rev. Lett. \textbf{99}, 201301 (2007);
P. M. Alsing, J. P. Dowling and G. J. Milburn, Phys. Rev. Lett. \textbf{94},
220401 (2005).

\bibitem {Bogoliubov}N. N. Bogoliubov, Zh. Eksperim. i Teor. Fiz. \textbf{34},
73 (1958) [English transl.: Sovet Phys. - JETP \textbf{7}, 51 (1958)].

\bibitem {Birrel}N. D. Birrel and P. C. W. Davies, \textit{Quantum Fields in
Curved Space} (Cambridge: Cambridge University Press, 1982); R. d'Inverno,
\textit{Introducting Einstein's Relativity }(Oxford: Clarendon Press, 1992).

\bibitem {Law}C. K. Law, Phys. Rev. A \textbf{49}, 433 (1994).

\bibitem {Soff1}G. Schaller, R. Sch\"{u}tzhold, G. Plunien, and G. Soff, Phys.
Rev. A \textbf{66}, 023812 (2002). 

\bibitem {Soff2}G. Plunien, R. Sch\"{u}tzhold, and G. Soff, Phys. Rev. Lett.
\textbf{84}, 1882 (2000); V. V. Dodonov and A. B. Klimov, Phys. Rev. A
\textbf{53}, 2664 (1996).

\textbf{Figure Captions}

Fig. 1 The mean number of particles created in the fundamental mode,
$\mathcal{N}_{\mathbf{1}}$, plotted against $\gamma a_{p}$, for a fixed time
$\tau_{p}=0.1$, setting $\chi=0$.

Fig. 2 The mean number of particles created in the $\mathbf{k}$th mode,
$\mathcal{N}_{\mathbf{k}}$, plotted against $\gamma a_{p}$, for a fixed time
$\tau_{p}=0.1/\mathbf{n}$, setting $\chi=0$.
\end{thebibliography}
\end{document}